\documentclass[preprint,showpacs]{revtex4}
\usepackage{epsfig}
\usepackage{graphicx}
\usepackage{bm}
\usepackage{amsmath} 
\newcommand{\rd}{\textrm{d}}
\begin{document}
%\noshowpacs
%{\flushright{\today}}

\title{Production and sequential decay of charmed hyperons }
\date{\today}
\author{G\"oran F\"aldt}\email{goran.faldt@physics.uu.se}  
\affiliation{ Department of physics and astronomy, \
Uppsala University,
 Box 516, S-751 20 Uppsala,Sweden }

%\date{Received: \today / Revised version: date}

\begin{abstract}
We investigate production and  decay of the $\Lambda_c^+ $ hyperon. The production considered 
is  through the $e^+e^-$ annihilation channel, $e^+e^-\rightarrow\Lambda_c^+ \bar{\Lambda}_c^-$, with 
summation over  the 
$\bar{\Lambda}_c^- $ anti-hyperon spin directions.  It is in this situation that the $\Lambda_c^+ $ decay  chain is identified. 
Two kinds of sequential decays are studied. The first one is the doubly weak decay 
 $B_1\rightarrow B_2 M_2$,
followed by $B_2 \rightarrow B_3 M_3$. The other one is the mixed weak-electromagnetic 
decay  $B_1\rightarrow B_2 M_2$, 
followed by $B_2 \rightarrow B_3 \gamma$. In both schemes  $B$ denotes baryons and $M$ mesons. 
We should also mention that the initial state of the  $\Lambda_c^+ $ hyperon  is polarized.

\end{abstract}

% \begin{keyword}
%Hadron production in e$^-$e$^+$ interactions, Hadronic decays
%\end{keyword}

%\noshowpacs
\maketitle
%

%
%%%%%%%%%%%%%%%%%%%%%%%%%%%
%
\section{Introduction}\label{ett}

We shall investigate properties of certain sequential decays of the $\Lambda_c^+$ hyperon, 
but in order to do so we first   need to produce them. To this end we consider the reaction 
$e^+ e^- \rightarrow\Lambda_c^+ \bar{\Lambda}_c^-$, which  is analyzed in detail in 
Refs.\cite{GF2,GF3}. In order to describe such an annihilation  process two hadronic form factors 
are needed. They can be parametrized by two parameters, $\alpha$ and $\Delta\Phi$, 
with $-1\leq \alpha\leq 1$. 
For their precise definitions we refer to Ref.~\cite{GF3}. 

The general cross-section distribution of this annihilation reaction depends on six structure functions which 
themselves are functions 
of $\alpha$, $\Delta\Phi$, and $\theta$, the scattering angle. 
In our application, however, we sum over the decay products of the anti-hyperon $\bar{\Lambda}_c^- $, 
but identify the decay chain of the hyperon $\Lambda_c^+$, so called single tag events.
 In this simplified case  only two  structure functions 
are relevant,
\begin{eqnarray}
{\cal{R}} &=& 1 +\alpha  \cos^2\!\theta, \label{DefR}\\
  {\cal{S}} &=& \sqrt{1-\alpha^2}\sin\theta\cos\theta\sin(\Delta\Phi). \label{DefS}
\end{eqnarray}

The scattering  distribution function for the $\Lambda_c^+$ hyperon production 
becomes, according to Refs.\cite{GF3}, 
proportional to
 \begin{equation}
	W(\mathbf{n}) =   {\cal{R}}  
	    + {\cal{S}}\, \mathbf{N}\cdot \mathbf{n},
	\label{ccdef}
\end{equation}
where $\mathbf{n}$ is the direction of the hyperon spin vector in the hyperon rest system,
 $\mathbf{N}$ the normal to the scattering plane,
\begin{equation}
 \mathbf{N}=\frac{1}{\sin\theta} \,  \hat{\mathbf{p}}\times \hat{\mathbf{k}},\label{Ndef}
\end{equation}
and $\cos\theta=\hat{\mathbf{p}}\cdot \hat{\mathbf{k}}$. The momenta $\mathbf{k}$ and $\mathbf{p}$
are the relative momenta in the initial and final states, in the center of momentum (c.m.) system. 
The meaning of the spin vector $ \mathbf{n}$ is explained in Ref.\cite{BJ}

From Eq.(\ref{ccdef}) we deduce for the spin distribution function, 
\begin{equation}
S(\mathbf{P}) =   1 + \mathbf{P}  \cdot \mathbf{n} \label{Hdef}
\end{equation}
and $\mathbf{P}$ the hyperon polarization,
\begin{equation}
\mathbf{P} =   ( {\cal{S}}/ {\cal{R}} )  \mathbf{N},
	\label{Pdef}
\end{equation}
  subject to  the restriction $|\mathbf{P}|\leq 1$. 
For an unpolarized 
initial state hyperon $\mathbf{P}=0$. 

%
%
%
%%%%%%%%%%%%%%%%%%%%%%%%%%%
%
\section{Weak hyperon decays}\label{HD}

The weak hyperon decay $c\rightarrow d\pi$, of which $\Lambda\rightarrow p\pi^-$ is an example,
 is described by two amplitudes, one S-wave
and one P-wave amplitude. The decay distribution is commonly described by  
three  parameters, denoted $\alpha\beta\gamma$. They are not independent but fulfill the relation 
\begin{equation}
	\alpha^2 + \beta^2 +\gamma^2=1.
\end{equation}
The parametrization of this hyperon decay is discussed in detail in 
Ref.\cite{Okun} and also  in Ref.\cite{GF3}.

We denote by  $G_c(c,d)$ the  distribution function describing the weak hyperon decay $c\rightarrow d\pi$,
given the spin vectors $\mathbf{n}_c\textbf{}$ and $\mathbf{n}_d$,
\begin{equation}
	G_c(c, d) = 1+\alpha_c \mathbf{n}_c\cdot \mathbf{l}_d
	  +\alpha_c \mathbf{n}_d\cdot \mathbf{l}_d
		+\mathbf{n}_c\cdot \mathbf{L}_c(\mathbf{n}_d,\mathbf{l}_d ),
		\label{GcDef}
\end{equation}	
with
\begin{equation}
	\mathbf{L}_c(\mathbf{n}_d, \mathbf{l}_d)=\,\gamma_c \mathbf{n}_d
	+\bigg[(1-\gamma_c)\mathbf{n}_d\cdot \mathbf{l}_d\bigg]\, \mathbf{l}_d
	+\beta_c  \mathbf{n}_d\times \mathbf{l}_d .
\end{equation}
The vector $\mathbf{l}_d$ is a unit vector in the direction of motion
of the decay baryon $d$ in the rest system of baryon $c$. The indices on the $\alpha\beta\gamma$ 
parameters remind us they characterize  hyperon $c$.

Since the spin of baryon $d$ is often not measured, the relevant decay density is obtained by 
 averaging over the spin directions $\mathbf{n}_d$,
\begin{align}
  W_c(\mathbf{n}_c; \mathbf{l}_d) =&\, {\bigg\langle} 
	G_c(c,d) {\bigg\rangle}_d \nonumber \\
	=&\, U_c+  \mathbf{n}_c\cdot \mathbf{V}_c,\label{Gcd}
\end{align}
with
\begin{equation}
		U_c=1 , \qquad  \mathbf{V}_c= \alpha_c \mathbf{l}_d. \label{UVc}
\end{equation}

Hyperons we study are produced in some reaction, and their states are described by some 
spin distribution function, Eq.(\ref{Hdef}), 
\begin{equation}
	S_c(\mathbf{P}_c)= 1+\mathbf{P}_c\cdot \mathbf{n}_c. \label{So1}
\end{equation}

The final-state distribution in a  production reaction followed by decay is obtained by 
a folding, pertaining to the intermediate and final hyperon spin directions $\mathbf{n}_c$ and $\mathbf{n}_d$,
\begin{align}
	 W_c(\mathbf{P}_c;\mathbf{l}_d)= &\,{\bigg\langle} S_c(\mathbf{P}_c)
	G_c(c,d) {\bigg\rangle}_{cd}
	\nonumber\\
	= &\,1+ \mathbf{P}_c\cdot \mathbf{V}_c,\label{Waceq}
\end{align}
where $\mathbf{V}_c=\alpha_c \mathbf{l}_d$, from Eq.(\ref{UVc}).

The folding over intermediate spin directions follows the prescription of Ref.\cite{GF2},
\begin{equation}
\big{\langle } 1\big{\rangle }_{\mathbf{n}}   =1, \quad 
\big{\langle }  \mathbf{n} \big{\rangle }_{\mathbf{n}}   =0, \quad
\big{\langle }  \mathbf{n}\cdot \mathbf{k}  \mathbf{n}\cdot \mathbf{l}  \big{\rangle }_{\mathbf{n}}   
=\mathbf{k}\cdot \mathbf{l} .\label{Defaverage}
 \end{equation}
 
 From Eq.(\ref{Waceq}) it is clear that if the polarization is known the asymmetry parameter 
 $\alpha_c$ can be measured, but not the $\beta_c$ or $\gamma_c$ parameters. For that to be possible
 we must measure the polarization of the decay baryon $d$. If hyperon $c$  is produced within a $c\bar{c}$ 
 pair in $e^+e^-$ annihilation then the polarization can be determined from the cross-section distribution.

%
%
%%%%%%%%%%%%%%%%%%%%%%%%%%%%%%%%%%%%%%%%%%%%%%%%
%
\section{Electromagnetic hyperon transitions}\label{EM3}

Electromagnetic transitions such as $\Sigma^0\rightarrow\Lambda\gamma$ and 
$\Xi^0\rightarrow\Lambda\gamma$ can also be studied in $\Lambda_c^+$ decays.

An electromagnetic transition $c\rightarrow d\gamma$ is described by a transition distribution function  
similar to that of the weak decay, Eq.(\ref{GcDef}). 
However, the  special feature of the electromagnetic interaction is the photon helicity which takes 
only two values, $ \lambda_\gamma=\pm 1$.

The electromagnetic transition distribution function  corresponding to Eq.(\ref{GcDef}) is
\begin{equation}
	G_{\gamma}(cd; \lambda_\gamma)= (1-
	\mathbf{n}_{c}\cdot \mathbf{l}_d \mathbf{l}_d\cdot \mathbf{n}_{d})
	 -\lambda_\gamma (\mathbf{n}_{c}\cdot \mathbf{l}_d 
		-\mathbf{n}_{d}\cdot \mathbf{l}_d), 
	\label{Eltrans}
\end{equation}	
where $\mathbf{l}_d$ is a unit vector in the direction of motion of hyperon $d$ in the 
rest system of hyperon $c$.

Averaging over  photon polarizations the transition distribution takes a very simple form,
\begin{equation}
	G_\gamma(c,d)= 1-
	\mathbf{n}_{c}\cdot {\mathbf{l}}_{d} {\mathbf{l}}_{d}\cdot \mathbf{n}_{d}.
	\label{Photav}
\end{equation}	

We notice that when both hadron spins are parallel or anti-parallel to the photon
momentum, then the transition probability vanishes, a property of angular-momentum conservation.
We also notice that expression (\ref{Photav}) cannot be written in the 
$\alpha\beta\gamma$ representation of Eq.(\ref{GcDef}). 
%
%
%
%
%%%%%%%%%%%%%%%%%%%%%%%%%%%%%%%%%%%%%%%%%%%%%%%%%
\section{Two-step weak hyperon decay}\label{HDbis}

Now, we apply the above technique to hyperons decaying in two steps,
such as $b\rightarrow c\rightarrow d$, accompanied by pions. 
An example of this decay mode is $ \Lambda_c^+\rightarrow\Lambda\pi^+$ followed by $\Lambda\rightarrow p\pi^-$. 

We denote by  $G_b(b,c)$ the distribution function  describing the hyperon decay $b\rightarrow c\pi$
pertaining to
spin vectors $\mathbf{n}_b\textbf{}$ and $\mathbf{n}_c$,
\begin{equation}
	G_b(b,c) = 1+\alpha_b \mathbf{n}_b\cdot \mathbf{l}_c
	  +\alpha_b \mathbf{n}_c\cdot \mathbf{l}_c
		+\mathbf{n}_b\cdot \mathbf{L}_b(\mathbf{n}_c,\mathbf{l}_c ),
		\label{GbDef}
\end{equation}	
with
\begin{equation}
	\mathbf{L}_b(\mathbf{n}_c, \mathbf{l}_c)=\,\gamma_b \mathbf{n}_c
	+\bigg[(1-\gamma_b)\mathbf{n}_c\cdot \mathbf{l}_c\bigg]\, \mathbf{l}_c
	+\beta_b  \mathbf{n}_c\times \mathbf{l}_c .
\end{equation}
The vector $\mathbf{l}_c$ is a unit vector in the direction of motion
of baryon $c$ in the rest system of baryon $b$.

Folding together the distribution functions $G_b(b,c)$ and $G_c(c,d)$, averaging over spin vectors
 $\mathbf{n}_c$ and $\mathbf{n}_d$ following the folding prescription (\ref{Defaverage}), we get
the decay density distribution function
\begin{align}
 W_b(\mathbf{n}_b;  \mathbf{l}_c, \mathbf{l}_d) =&\,
 {\bigg\langle} G_b(b,c)
	G_c(c,d) {\bigg\rangle}_{cd}  \nonumber \\ =&\, U_b+  \mathbf{n}_b\cdot \mathbf{V}_b, \label{GUVav}
\end{align}
with
\begin{align}
		U_b=&1 +\alpha_b\alpha_c \mathbf{l}_c \cdot \mathbf{l}_d, \label{BUU} \\
		  \mathbf{V}_b=&\alpha_b \mathbf{l}_c +\alpha_c\mathbf{L}_b(\mathbf{l}_d,\mathbf{l}_c ).
		  \label{BUV}
\end{align}

The result is interesting. In many cases  the asymmetry parameter $\alpha_c$ for the $c$
hyperon
and the polarization $\mathbf{P}_b$ for the initial-state $b$ hyperon are known. 
Then, just as in the single-step case of  Eq.(\ref{So1}), 
the initial state  is  described by a spin distribution function 
\begin{equation}
	S_b(\mathbf{P}_b)= 1+\mathbf{P}_b\cdot \mathbf{n}_b.  \label{So2}
\end{equation}
 For the decay distribution of a polarized hyperon, 
we obtain
\begin{align}
 W_b(\mathbf{P}_b;  \mathbf{l}_c, \mathbf{l}_d) =&\,
 {\bigg\langle}  S_b( \mathbf{P}_b) G_b(b,c)
	G_c(c,d) {\bigg\rangle}_{bcd}  \nonumber \\ =&\, U_b+  \mathbf{P}_b\cdot \mathbf{V}_b. \label{GBbav}
\end{align}  
This is equivalent to making the replacement $\mathbf{n}_b\rightarrow\mathbf{P}_b$
in Eq.(\ref{GUVav}). 

We conclude that by determining $U_b$ and $\mathbf{V}_b$ of Eqs.(\ref{BUU}) and 
(\ref{BUV}), we should be able to determine all three decay parameters $\alpha_b$, $\beta_b$, and $\gamma_b$,
for the $b$ hyperon, and $\alpha_c$ for the $c$ hyperon. 

It is now clear how to get  the cross-section distribution for 
 production of $\Lambda^+_c$ in $e^+e^-$ annihilation and its
subsequent decay  $\Lambda^+_c\rightarrow\Lambda \pi^+ $ and $\Lambda\rightarrow p \pi^-$.
Starting from the expressions for the scattering distribution function, 
Eq.(\ref{ccdef}), and the polarization, Eq.(\ref{Pdef}). 
we obtain
\begin{equation}
	\rd \sigma \propto \bigg[ {\cal{R}} U_{\Lambda_c} + {\cal{S}}
	\mathbf{N}\cdot \mathbf{V}_{\Lambda_c} \bigg] \rd\Omega_{\Lambda_c} 
	\rd\Omega_{\Lambda} \rd\Omega_p , \label{Twocc}
\end{equation} 
with $\mathbf{N}$, Eq.(\ref{Ndef}), the normal to the scattering plane. The functions $ \cal{R}$ and 
$ \cal{S}$ are defined in Eqs.(\ref{DefR}) and (\ref{DefS}) and depend among other things on the 
$\Lambda^+_c$ scattering angle $\theta$ (=$\theta_{\Lambda_c}$). 
In Eqs.(\ref{BUU}) and (\ref{BUV})  indices are interpreted as; $b=\Lambda^+_c$, $c=\Lambda$, 
$d=p$.

When integrating over the decay angles $\Omega_{\Lambda}$ and $\Omega_p $  in Eq.(\ref{Twocc}) 
we observe that the term involving the polarization $\mathbf{N}\cdot \mathbf{V}_{\Lambda_c}$ 
vanishes, as does the term involving the angular dependent part of $U_{\Lambda_c}$. This results is the 
cross-section distribution
\begin{equation}
	\rd \sigma \propto \bigg[1 +\alpha  \cos^2\!\theta_{\Lambda_c}\bigg] \rd\Omega_{\Lambda_c}  ,
\end{equation} 
describing the annihilation reaction 
$e^+ e^- \rightarrow\Lambda_c^+ \bar{\Lambda}_c^-$.

It is more interesting to perform a partial integration. Let us integrate over the angles 
 $\Omega_{\Lambda}$ and $\Omega_p $ keeping $\cos\theta_{\Lambda p}$ of 
\begin{equation}
	\cos\theta_{\Lambda p}=\mathbf{l}_\Lambda  \cdot \mathbf{l}_p 
\end{equation}
constant. Also in this case does the contribution involving the polarization vanish. We are left with
\begin{equation}
	\rd \sigma \propto \bigg[1 +\alpha  \cos^2\!\theta_{\Lambda_c}\bigg]
	\bigg[1+\alpha_{\Lambda_c}\alpha_\Lambda \cos\theta_{\Lambda p} \bigg]
	\rd(\cos\theta_{\Lambda_c}) \rd(\cos\theta_{\Lambda p}) .	\label{Two_angles}
\end{equation}

The cross-section distribution of Eq.(\ref{Twocc})  applies also to the decay chain, 
 $\Lambda^+_c\rightarrow\Sigma^+ \pi^0 $ and $\Sigma^+\rightarrow p \pi^0 $, with the corresponding 
identification of indices $b,$ $c$, and $d$.
%
%
%%========================================
%

\section{Differential distributions}

The cross-section distribution (\ref{Twocc}) is a function of two unit vectors 
$\mathbf{l}_1=\mathbf{l}_{\Lambda}$, the direction of motion of the Lambda 
hyperon in the rest system of the charmed-Lambda hyperon, and 
$\mathbf{l}_2=\mathbf{l}_{p}$ the direction of motion of the proton 
 in the rest system of the Lambda hyperon. In order to handle these
vectors we need a common coordinate system which we define as follows.

The scattering plane  of the reaction 
$e^+e^-\rightarrow\Lambda_c\bar{\Lambda}_c$ is spanned by the unit vectors 
 $\hat{\mathbf{p}}=\mathbf{l}_{\Lambda_c}$ and $\hat{\mathbf{k}}=\mathbf{l}_{e^+}$, as 
measured in  the c.m.\ system. 
We assume the scattering to be to the left, with scattering angle $\theta\geq 0$. 
If the scattering is to the right we rotate such an event $180^{\circ}$ 
around the $\mathbf{k}$-axis, 
so that the scattering appears to be to the left. The scattering plane makes up 
the $xz$-plane, with the $y$-axis  along the normal to the scattering plane. 
We choose a right-handed coordinate system with basis vectors 
	\begin{eqnarray}
	\mathbf{e}_z  &=&  \hat{\mathbf{p}}, \label{zunity}\\
	\mathbf{e}_y  &=& \frac{1}{\sin\theta } ( \hat{\mathbf{p}}\times \hat{\mathbf{k}} ) ,\label{yunity} \\
	\mathbf{e}_x  &=& \frac{1}{\sin\theta } (\hat{\mathbf{p}}\times \hat{\mathbf{k}} ) 
	 \times\hat{\mathbf{p}}.\label{xunity}
\end{eqnarray}
Expressed in terms of them the initial-state momentum 
\begin{equation}
	\hat{\mathbf{k}}= \sin\theta\,  \mathbf{e}_x +\cos\theta\,	\mathbf{e}_z  .
\end{equation}

This coordinate system is used for defining the directional angles of 
the  Lambda and the  proton. The  directional angles of the Lambda hyperon 
in the charmed-Lambda hyperon rest system are,
\begin{equation}
	\mathbf{l}_1=(\cos \phi_1 \sin \theta_1,  \sin \phi_1 \sin \theta_1, \cos \theta_1),
\end{equation}
whereas the  directional angles of the proton in the Lambda hyperon rest system are 
\begin{equation}
	\mathbf{l}_2=(\cos \phi_2 \sin \theta_2,  \sin \phi_2 \sin \theta_2, \cos \theta_2).
\end{equation}

An event of the reaction 
$e^+e^-\rightarrow \bar{\Lambda}_c \Lambda_c(\rightarrow \Lambda(\rightarrow p\pi)\pi)$
 is specified by the five dimensional vector 
${\boldsymbol{\xi}}=(\theta,\Omega_{1},\Omega_{2})$, and the differential-cross-section distribution as
summarized  by Eq.(\ref{Twocc}) reads,
\[
{\rd\sigma}\propto {\cal{W}}({\boldsymbol{\xi}})\ {\rd\!\cos\theta\ \rd\Omega_{1}\rd\Omega_{2}}.
\]
At the moment, we are not interested in  absolute normalizations. 
The differential-distribution function ${\cal{W}}({\boldsymbol{\xi}})$ 
is obtained from Eqs.(\ref{DefR}, \ref{DefS},\ref{BUU}, \ref{BUV},  \ref{Twocc}) and can be expressed as,
\begin{equation}
\begin{split}
{\cal{W}}({\boldsymbol{\xi}})=&\ {\cal{F}}_0({\boldsymbol{\xi}})+{{{\alpha}}}{\cal{F}}_1({\boldsymbol{\xi}})
   +\alpha_1\alpha_2\ {\bigg(} {\cal{F}}_2({\boldsymbol{\xi}})+{{{\alpha}}}{\cal{F}}_3({\boldsymbol{\xi}})\bigg)\\
	+&\  \sqrt{1-{{\alpha}}^2}\cos({{\Delta\Phi}})\, \bigg( {\cal{F}}_7({\boldsymbol{\xi}}) 
	    +  {{\alpha_{1}}}{\cal{F}}_4 ({\boldsymbol{\xi}}) + \beta_1 {\cal{F}}_6({\boldsymbol{\xi}})\\
	+&\       \gamma_1 \left( {\cal{F}}_5({\boldsymbol{\xi}})
	- {\cal{F}}_7({\boldsymbol{\xi}})\right)\bigg) ,  \label{eqn:pdf}
\end{split}
\end{equation}
%%%%%%%%%%%%%%%%%%%%%%%%%%
using a set of eight angular 
functions ${\cal{F}}_k({\boldsymbol{\xi}})$ defined as:
\begin{align}
	{\cal{F}}_0({\boldsymbol{\xi}}) =&1, \nonumber\\
	{\cal{F}}_1({\boldsymbol{\xi}}) =&{\cos^2\!\theta},\nonumber\\
	{\cal{F}}_2({\boldsymbol{\xi}}) =&\sin\theta_1\sin\theta_2\cos(\phi_1-\phi_2)+
              \cos\theta_1\cos\theta_2 ,\nonumber\\
	{\cal{F}}_3({\boldsymbol{\xi}}) =& {\cos^2\!\theta}\ {\cal{F}}_2({\boldsymbol{\xi}}),  \nonumber\\
	{\cal{F}}_4({\boldsymbol{\xi}}) =& \sin\theta \cos\theta\sin\theta_1 \sin\phi_1	,\nonumber\\
	{\cal{F}}_5({\boldsymbol{\xi}}) =& \sin\theta \cos\theta\sin\theta_2 \sin\phi_2 ,	\nonumber\\
	{\cal{F}}_6({\boldsymbol{\xi}}) =&\sin\theta \cos\theta\  (\cos\theta_2 \sin\theta_1\cos\phi_1 \nonumber\\
	    &\qquad \qquad \qquad -\cos\theta_1 \sin\theta_2\cos\phi_2 ) , \nonumber\\
{\cal{F}}_7({\boldsymbol{\xi}}) =&\sin\theta \cos\theta\sin\theta_1\sin\phi_1 \ {\cal{F}}_2({\boldsymbol{\xi}}).
\end{align}

The differential distribution of Eq.~(\ref{eqn:pdf}) involves two parameters related to the $e^+e^-\to\Lambda_c
\bar{\Lambda}_c$ reaction that can be determined by data: the ratio of form factors $\alpha$,  and the relative phase of form factors $\Delta\Phi$. 
In addition, the distribution function ${\cal{W}}({\boldsymbol{\xi}})$ depends on the weak-decay parameters 
$\alpha_1\beta_1\gamma_1$ of the charmed-hyperon decay $\Lambda_c\rightarrow \Lambda\pi$,
and on the weak-decay parameters 
$\alpha_2\beta_2\gamma_2$ of the hyperon decay $\Lambda\rightarrow p\pi^-$.
However, the dependency on $\beta_2$ and $\gamma_2$ drops out. Similarly, integrating over  $\rd\Omega_{2}$
we get
\begin{align}
{\rd\sigma}\propto \bigg[ 1+&\alpha \, {\cos^2\!\theta} \nonumber \\
  +&\alpha_1\sqrt{1-{{\alpha}}^2}\cos({{\Delta\Phi}})
 \sin\theta\cos\theta\sin\theta_1 \sin\phi_1  \bigg]\, \rd\Omega \, \rd\Omega_{1} ,
\end{align}
where  now  the dependency on $\beta_1$ and $\gamma_1$ also drops out.
The last term in this equation originates with the scalar $\mathbf{P}_{\Lambda_c}\cdot \mathbf{N}$.
The charmed-hypern polarization vanishes at $\theta=0^\circ$, $90^\circ$ and
$180^\circ$. 
 
The distributions presented here will hopefully be of value in the analysis of    BESIII data.

%
%
%
%%%%%%%%%%%%%%%%%%%%%%%%%%%%%%%%%%%%%%%%%%%%%%%%%
\section{Mixed weak-electromagnetic hyperon decay}\label{HDbisII}

Now, we extend the formalism to hyperons decaying in two steps, with one being electromagnetic. 
An example of such a decay chain is  
$ \Lambda_c^+\rightarrow\Sigma^0\pi^+$ followed by $\Sigma^0\rightarrow \Lambda \gamma$. 
As before we employ indices $b$, $c$, and $d$ for variables belonging to 
$\Lambda_c^+$, $\Sigma^0$, and $\Lambda$.

The distribution functions for the weak and electromagnetic transitions are given 
in Eqs.(\ref{GbDef}) and (\ref{Photav}),
\begin{align}
	G_b(b, c) &= 1+\alpha_b \mathbf{n}_b\cdot \mathbf{l}_c
	  +\alpha_b \mathbf{n}_c\cdot \mathbf{l}_c
		+\mathbf{n}_b\cdot \mathbf{L}_b(\mathbf{n}_c,\mathbf{l}_c ), \\
			G_\gamma (c,d) &= 1-
	\mathbf{n}_c\cdot \mathbf{l}_d \mathbf{l}_d\cdot \mathbf{n}_d.
\end{align}

Performing a folding of the product of the distribution functions
 $G_b(b,c)$ and $G_\gamma (c,d)$, { \slshape i.e.}\ 
averaging over spin vectors
 $\mathbf{n}_c$ and $\mathbf{n}_d$ following the folding prescription (\ref{Defaverage}), we get 
\begin{align}
 W_b(\mathbf{n}_b;  \mathbf{l}_c, \mathbf{l}_d) =&\,
 {\bigg\langle} G_b(b,c)
	G_\gamma (c,d) {\bigg\rangle}_{cd}  \nonumber \\ =&\, U_b+  \mathbf{n}_b\cdot \mathbf{V}_b, \label{Gbav}
\end{align}
with
\begin{equation}
		U_b=1 , \qquad  \mathbf{V}_b= \alpha_b \mathbf{l}_c. \label{UVEc}
\end{equation}

These expressions for $U_b$ and $\mathbf{V}_b$  are noteworthy. They are in fact the same as those of a one-step $b\rightarrow c\pi$ decay,  
Eq.(\ref{UVc}). Hence, the electromagnetic decay does not add any structure,  Eqs.(\ref{UVEc}) 
are independent of $\mathbf{l}_d$.

The initial state spin distribution function for hyperon $b$ produced in 
$e^+e^-$ annihilation is as above, Eq.(\ref{So2}),  
\begin{equation}
	S_b(\mathbf{P}_b)= 1+\mathbf{P}_b\cdot \mathbf{n}_b, \label{So3}
\end{equation}
Folding this distribution function with the decay distribution function of Eq.(\ref{Gbav}), we obtain
\begin{align}
 W_b(\mathbf{P}_b;  \mathbf{l}_c, \mathbf{l}_d) =&\,
 {\bigg\langle}  S_b( \mathbf{P}_b) G_b(b,c)
	G_\gamma (c,d) {\bigg\rangle}_{bcd}  \nonumber \\ =&\, U_b+  \mathbf{P}_b\cdot \mathbf{V}_b. \label{GBW}
\end{align}  
As noted earlier this is equivalent to making the replacement $\mathbf{n}_b\rightarrow\mathbf{P}_b$
in Eq.(\ref{Gbav}). We also notice if we manage to determine $U_b$ and $\mathbf{V}_b$ of Eqs.(\ref{GBW}), 
the only parameter that  can be fixed is  $\alpha_b$, a meager return. 

The expression for the cross-section distribution for $\Lambda^+_c$ production and 
subsequent decays $\Lambda^+_c\rightarrow\Sigma^0 \pi^+ $ and $\Sigma^0\rightarrow \Lambda \gamma $ is
\begin{equation}
	\rd \sigma \propto \bigg[ {\cal{R}} U_{\Lambda_c}+ {\cal{S}} \mathbf{N}\cdot \mathbf{V}_{\Lambda_c}\bigg] \rd\Omega_{\Lambda_c} 
	\rd\Omega_{\Sigma} \rd\Omega_{\Lambda} , 
\end{equation}
with $\mathbf{N}$, Eq.(\ref{Ndef}), the normal to the scattering plane, and 
$U_{\Lambda_c}=1$,   $\mathbf{V}_{\Lambda_c}=\alpha_{\Lambda_c } \mathbf{l}_{\Sigma}$, 
from Eq.(\ref{GBW}). The functions $ \cal{R}$ and 
$ \cal{S}$ are defined in Eqs.(\ref{DefR}) and (\ref{DefS}) and depend among other things on the 
$\Lambda^+_c$ scattering angle $\theta$. 

Finally, we mention that it is possible to expand the $\Lambda^+_c$ decay chain  by adding  the decay 
$\Lambda\rightarrow p\pi^-$. 
%
%%%%%%%%%%%%%%%%%%%%%%%%%%%%%%%%%%%%%%%%%%%%%%%%%%%%%%%%%%%%%%%%%%%%%%%%%%%%%%%%%%%%%%%%%%%

\section*{Acknowledgments}
Thanks to Andrzej Kupsc and Karin Sch\"onning for valuable discussions and suggestions. 
%

%
%%%%%%%%%%%%%%%%%%%%%%%%%%%%%%%%%%%%
\section*{Appendix}
In this Appendix we detail the angular integration leading to Eq.(\ref{Two_angles}).

Consider two unit vectors $\mathbf{l}_c$ and $\mathbf{l}_d$. We want to integrate over the 
angles $\Omega_c$ and $\Omega_d$ keeping $\cos\theta_{cd}=\mathbf{l}_c\cdot\mathbf{l}_d$ fixed.
To this end we put the vectors in the $xy$-plane of the coordinate system O',
\begin{eqnarray}
	\mathbf{l}_c&=& (1,0,0),\\
	\mathbf{l}_d&=& (\cos\theta_{cd},\sin\theta_{cd},0), \\
	\mathbf{l}_c\times\mathbf{l}_d&=& (0,0,\sin\theta_{cd}).
\end{eqnarray}
We then rotate the coordinate system O' with respect to the space fixed coordinate system O, 
where the normal to the scattering plane is along the $Z$-direction. The rotation matrix which
transforms the column vector $\bar{r}_b$ in O' into the column vector $\bar{r}_s$ in O is the matrix 
\begin{equation}
	R^{-1}(\alpha, \beta, \gamma )=\left( \begin{array}{lll} \cos \alpha \cos \beta\cos \gamma  
	  - \sin \alpha \sin \gamma  &  \cos \gamma \cos \beta \sin \alpha +\sin \gamma\cos \alpha
		& -\sin \beta        \cos \gamma                      \\
	    -\sin \gamma \cos \beta \cos \alpha   -\cos \gamma \sin \alpha   &  -\sin \gamma \cos \beta \sin \alpha 
			+\cos \gamma \cos \alpha                 & \sin \beta \sin \gamma \\
			 \cos \alpha \sin \beta  &  \sin \gamma \sin\beta  & \cos \beta 
	\end{array} \right).
\end{equation}
with $\bar{r}_s=R^{-1}(\alpha, \beta, \gamma )\bar{r}_b$ and $\alpha\beta\gamma$ the Euler angles.

The angular integrations can be expressed in terms of the Euler angles, as
\begin{equation}
	\rd\Omega_c \, \rd\Omega_d=\rd(\cos\theta_{cd})\,\rd\alpha\, \rd(\cos \beta)\,  \rd\gamma.
\end{equation}
The expression to be integrated, Eq.(\ref{BUV}),  reads
\begin{align}
		U_b+ \mathbf{P}\cdot\mathbf{V}_b =&1 +\alpha_b\alpha_c \mathbf{l}_c \cdot \mathbf{l}_d
		  +\alpha_b\mathbf{P}\cdot \mathbf{l}_c +\gamma_b\mathbf{P}\cdot \mathbf{l}_d\nonumber \\
		& 
	+(1-\gamma_b)\mathbf{P}\cdot  \mathbf{l}_c\mathbf{l}_d\cdot \mathbf{l}_c
	+\beta_b \mathbf{P}\cdot ( \mathbf{l}_d\times \mathbf{l}_c) ,
\end{align}
with $\mathbf{P}$ along the $Z$-direction.

Now, we note that terms proportional to $\mathbf{P}\cdot \mathbf{l}_c$ or 
$\mathbf{P}\cdot \mathbf{l}_d$ vanish upon integration over angles $\alpha$ or $\gamma$.
Therefore,
\begin{align}
		& \int \rd\Omega_c\rd\Omega_d\,(U_b+ \mathbf{P}\cdot\mathbf{V}_b)  = \nonumber\\
		& \qquad=4\pi^2 \int\rd(\cos\theta_{cd})\rd (\cos\beta) \, 
		\bigg(1  +\alpha_b\alpha_c  \cos\theta_{cd} \nonumber\\
		&	\hspace{4cm}-\beta_b\alpha_c P\sin\theta_{cd}\cos\beta\bigg)&
						 \nonumber\\&
		\qquad=8\pi^2 
		\int \rd(\cos\theta_{cd})\, \bigg(1 +\alpha_b\alpha_c \cos\theta_{cd} \bigg).
\end{align}
This result leads to Eq.(\ref{Two_angles}).

%%%%%%%%%%%%%%%%%%%%%%%%%%%%%%%%%%%%%%%%%%%%%%%%%%%%%
%


\begin{thebibliography}{99}
\bibitem{GF2} G. F\"aldt, Eur.\ Phys.\ J.\ A  {\bf 52},  141 (2016).
% \bibitem{GF1} G. F\"aldt, Eur.\ Phys.\ J.\ A  {\bf 51},  74 (2015).
\bibitem{GF3} G. F\"aldt and A.Kupsc, Phys.\ Lett.\ B {\bf 772},  16 (2017).
%\bibitem{GF4} G. F\"aldt, to be published, (2017).
\bibitem{BJ} J.D.\ Bjorken and S.D.\ Drell, Relativistic Quantum Mechanics,
  Mc Graw-Hill, New York, 1964. 
\bibitem{Okun} L.B.\ Okun, Leptons and Quarks, North-Holland, Amsterdam,  1982.
%\bibitem{Ber} R.E.\ Behrends,  Phys.\ Rev. {\bf 111},  1691 (1958). 
%  \bibitem{Pil} H.\ Pilkuhn, \textit{Relativistic Particle Physics} (Springer-Verlag,
%  Berlin, 1979).

\end{thebibliography}
\end{document}